\documentclass[%
aip,
amsmath,amssymb,
reprint,%
]{revtex4-2}

\usepackage{graphicx}
\usepackage{dcolumn}
\usepackage{bm}

\usepackage[utf8]{inputenc}
\usepackage[T1]{fontenc}
\usepackage{mathptmx}
\usepackage{etoolbox}

\usepackage[hidelinks,unicode=true]{hyperref}
\hypersetup{colorlinks=true,
	linkcolor=blue,
	urlcolor=blue,
	citecolor=blue,
	pdfhighlight=/N
}

\usepackage{calc}

\makeatletter
\def\@email#1#2{%
	\endgroup
	\patchcmd{\titleblock@produce}
	{\frontmatter@RRAPformat}
	{\frontmatter@RRAPformat{\produce@RRAP{*#1\href{mailto:#2}{#2}}}\frontmatter@RRAPformat}
	{}{}
}%
\makeatother

\newcommand{\av}[1]{\langle {#1} \rangle}

\newcommand{\ds}{\ensuremath{\displaystyle}}
\newcommand{\dd}{\text{d}}
\newcommand{\dv}[2]{\frac{\dd #1}{\dd #2}}

\newcommand{\cS}{\text{S}}
\newcommand{\cSV}{\text{S}_\text{V}}
\newcommand{\cSP}{\text{S}_\text{P}}
\newcommand{\cE}{\text{E}}
\newcommand{\cEV}{\text{E}_\text{V}}
\newcommand{\cA}{\text{A}}
\newcommand{\cAV}{\text{A}_\text{V}}
\newcommand{\cI}{\text{I}}
\newcommand{\cIV}{\text{I}_\text{V}}
\newcommand{\cIP}{\text{I}_\text{P}}
\newcommand{\cR}{\text{R}}
\newcommand{\cRV}{\text{R}_\text{V}}
\newcommand{\cRP}{\text{R}_\text{P}}
\newcommand{\cD}{\text{D}}
\newcommand{\cDV}{\text{D}_\text{V}}
\newcommand{\cDP}{\text{D}_\text{P}}
\newcommand{\cV}{\text{V}}
\newcommand{\cP}{\text{P}}
\newcommand{\cX}{{\mathcal{Z}}}

\newcommand{\icS}{{S}}
\newcommand{\icSV}{{S}_\text{V}}
\newcommand{\icSP}{{S}_\text{P}}
\newcommand{\icE}{{E}}
\newcommand{\icEV}{{E}_\text{V}}
\newcommand{\icA}{{A}}
\newcommand{\icAV}{{A}_\text{V}}
\newcommand{\icI}{{I}}
\newcommand{\icIV}{{I}_\text{V}}
\newcommand{\icIP}{{I}_\text{P}}
\newcommand{\icR}{{R}}
\newcommand{\icRV}{{R}_\text{V}}
\newcommand{\icRP}{{R}_\text{P}}
\newcommand{\icD}{{D}}
\newcommand{\icDV}{{D}_\text{V}}
\newcommand{\icDP}{{D}_\text{P}}

\newcommand{\icP}{{P}}

\newcommand{\uppX}[1]{{\{\text{#1}\}}}

\newcommand{\lbcp}[1]{\lambda_{#1}}

\newcommand{\fupp}[1]{(#1)}

\newcommand{\ci}[1]{\ensuremath{#1^{\fupp{i}}}}
\newcommand{\cj}[1]{\ensuremath{#1^{\fupp{j}}}}

\newcommand{\ri}[2]{\ensuremath{#1_{#2}^{\fupp{i}}}}

\newcommand{\nni}{\ensuremath{n^{\fupp{i}}}}
\newcommand{\nnj}{\ensuremath{n^{\fupp{j}}}}
\newcommand{\nnl}{\ensuremath{n^{\fupp{l}}}}

\newcommand{\lbpar}{\ensuremath{\omega}}

\newcommand{\SM}{SM~\cite{supp}}
\newcommand{\rhoinf}{\rho_\text{inf}}

\newcommand{\rhod}{\rho_\text{d}}

\newcommand{\orcid}[1]{\hspace{0.2em}\href{https://orcid.org/#1}{\includegraphics[keepaspectratio,width=0.7em]{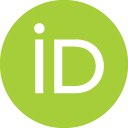}}}

\begin{document}
	
	
	\title[Effects of IFR and social contact matrices on vaccine prioritization strategies]{Effects of infection fatality ratio and social contact matrices on vaccine prioritization strategies}
	
	\author{Arthur Schulenburg\orcid{0000-0001-7548-7595}}
	\affiliation{%
		Departamento de Fisica, Universidade Federal de Viçosa,  Viçosa, Minas Gerais, 36570-900 Brazil %
	}
	
	\author{Wesley Cota\orcid{0000-0002-8582-1531}}
	\email{wesley@wcota.me}
	\altaffiliation[Also at ]{%
		Instituto de Medicina Tropical, Universidade de São Paulo, São Paulo,  05403-000 Brazil \& Departamento de Infectologia, Faculdade de Medicina de Botucatu, Universidade Estadual Paulista, Botucatu, São Paulo, 18618-687 Brazil
	}

	\affiliation{%
		Departamento de Fisica, Universidade Federal de Viçosa,  Viçosa, Minas Gerais, 36570-900 Brazil 
	}
	
	\author{Guilherme S. Costa\orcid{0000-0002-5019-0098}}
	\affiliation{%
		Departamento de Fisica, Universidade Federal de Viçosa,  Viçosa, Minas Gerais, 36570-900 Brazil 
	}

	\author{Silvio C. Ferreira\orcid{0000-0001-7159-2769}}
	\email{silviojr@ufv.br}
	\affiliation{%
		Departamento de Fisica, Universidade Federal de Viçosa,  Viçosa, Minas Gerais, 36570-900 Brazil 
	}
	\affiliation{%
	National Institute of Science and Technology for Complex Systems,  Rio de Janeiro, 22290-180 Brazil 
	}%
	
	\date{\today}
	
\begin{abstract}
Effective strategies of vaccine prioritization are essential to mitigate the impacts of severe infectious diseases. We investigate the role of infection fatality ratio (IFR) and social contact matrices on vaccination prioritization using a compartmental epidemic model fueled by real-world data of different diseases  and countries. Our study confirms that massive and early vaccination is extremely effective to reduce the disease fatality if the contagion is mitigated, but the effectiveness is increasingly reduced as vaccination beginning delays in an uncontrolled epidemiological scenario. The optimal and least effective prioritization strategies depend non-linearly on epidemiological variables. Regions of the epidemiological parameter space, in which prioritizing the most vulnerable population is more effective than the most contagious individuals, depend strongly on the IFR age profile being, for example, substantially broader for COVID-19 in comparison with seasonal influenza. Demographics and social contact matrices deform the phase diagrams but do not alter their qualitative shapes.

\noindent \\ \textbf{Accepted in Chaos: An Interdisciplinary Journal of Nonlinear Science}

\end{abstract}
	
	\maketitle
	
\begin{quotation}
Modeling of epidemic diseases allows the evaluation of possible strategies and their impacts in mitigating the threat of emerging infectious diseases. We investigate a mathematical model to {discern} the most effective vaccination strategies to reduce the fatality of infectious diseases. The method uses data from different countries (Brazil, Germany, and Uganda), social life scenarios (adopting or not social distancing), and other epidemiological parameters to fuel the computational simulations to analyze the most effective prioritization scheme. We report not only that early and massive vaccination is important, but also vaccinating first the most vulnerable individuals is the most effective {to reduce deaths due to the disease} in  highly infectious scenarios while prioritizing the most exposed population, who make more social contacts, can be more effective {in reducing the number of deaths} when the disease is spreading slowly. Determination of the most effective strategy is a multifactorial process that depends on disease specifics, such as the age profile of the disease fatality and vaccination efficacy, and non-biological features such as vaccination distribution, social contacts, and demographics.
\end{quotation}
	
\section{Introduction}

Densely connected and unequal societies impose enormous challenges for combating emerging infectious diseases and their potentially catastrophic consequences. The ongoing COVID-19 pandemic is an example that has reshaped the form of how people interact. Several and variable nonpharmacological interventions (NPIs) can be adopted across different places, as wearing of face masks, physical and social distancing, as well as more extreme ones such as lockdown, school closure, and traveling restrictions~\cite{Moore2021,Worby2020,Ventura2022,Sasikiran2020,Yacong2021}. While being clearly efficient to momentarily reduce the transmission and unburden {health} systems~\cite{Sasikiran2020,Yacong2021},  they are insufficient to restore the pre-pandemic lifestyle and avoid economic crashes caused by the disease~\cite{Vardavas2020}. Natural emergence of virus variants~\cite{Alpert2021, Alene2021, Planas2021, Naveca2021, Zhou2021,Campbell2021, Kupferschmidt2021}, fueled by negligent human behaviors, means that natural herd immunity by infections, in which individuals are immune and the susceptible pool is insufficient to a sustained transmission~\cite{Rohani2007}, is hard to be achieved~\cite{Sridhar2021}.

We have recently witnessed the development of vaccines for COVID-19 with unprecedented speed due to immense collaborative efforts, resources, and accumulated expertise from other viral infectious diseases~\cite{Heaton2020,Philip2021}. All vaccines approved for emergency use had a high potential to prevent severe cases after complete vaccination and immunization whose time depends on the type of vaccine. However, as usual for anti-viral vaccines, the capacity of COVID-19 vaccines to impede infections and mild symptoms is lower than their efficiency to reduce death and severe cases~\cite{Palacios2021,Yelin2021,Jara2021,Ranzani2021a,Frenck2021}. Indeed, while herd immunity is an important aim of massive vaccination, its main emergency function is to prevent severe cases which can result in deaths and serious sequelae~\cite{Goldstein2021,Fitzpatrick2021,Matrajt2020,Moret2021,Matrajt2021}.

The development of efficient vaccines is only the first challenge preceding massive immunization. Large-scale production and timely distribution, particularly to low-income economies, remain a major barrier to drastically reduce severe cases and to reach herd immunity~\cite{Li2021,Emanuel2020,Buckee2021}. Another great matter of concern, especially for high-income economies, is the low demand for vaccines by the {population}, which not rarely {refuses} to get their shots or to complete the immunization schedule. Therefore, given the finite capacity of vaccination, the logistic must be engineered to minimize damages~\cite{Castro2021,Goldstein2021,Fitzpatrick2021,Giordano2021,Bubar2021,Matrajt2021,Moret2021,Buckner2020,Molla2022,Milman2021}. 
Infection fatality ratio (IFR) is age- and illness-dependent due to, among other factors, cumulative comorbidities~\cite{PastorBarriuso2020,Ioannidis2020,Mallapaty2020,Barbosa2020,Castro2021a,Poletti2020,Levin2020}. While on the one hand, infectious diseases, and particularly COVID-19, present very high IFR on elderly in contrast with much lower values among kids and  newborns~\cite{Verity2020,Poletti2020,Levin2020,centers2020estimated}, on the other hand, younger population are socially more active and exposed to infections, being potentially the key vectors for contagion of the most vulnerable population~\cite{Davies2020,Johnson2021,Cevik2021}. Finally, but not least, the level of sustained transmission is also decisive for the degree of success in vaccination campaigns~\cite{Cevik2021,Bubar2021,Buckner2020,Molla2022,Epstein2021,Buckee2021}.

In the present work, we investigate the role of vaccination on an age-structured compartmental model~\cite{Rohani2007}, following a susceptible-exposed-asymptomatic-infected-recovered-deceased (SEAIRD) dynamics in different hypothetical scenarios, using social contact matrices~\cite{Prem2017} obtained for countries with very distinct population age distribution, namely Brazil, Germany, and Uganda. {The approach then consists fundamentally of coupling data on human behavior with epidemic and vaccination dynamics~\cite{Wang2016}.} Outcomes for age-dependent {IFR} estimated for COVID-19~\cite{Verity2020,Poletti2020,Levin2020} are compared with seasonal influenza~\cite{centers2020estimated}. The interplay between vaccine efficacy to prevent deaths after {complete} immunization and the time taken to acquire protection were also addressed. Epidemic scenarios representing different levels of NPIs were studied.  We have found that massive vaccination, even with modest protection against infections, is effective to reduce the disease fatality if adopted early and concomitantly with contagion mitigation, but losses effectiveness as the epidemic transmission becomes uncontrolled. Comparing different prioritization strategies, vaccinating the most vulnerable individuals first is the optimal strategy to reduce deaths for high transmission scenarios while distributing first shots to the most exposed ones can be better in lower transmission regimes. The region of the epidemiological parameters' space, in which  prioritization of the vulnerable population is the most effective strategy, {depends on the IFR and, thus, is disease dependent}. Demographics quantitatively change the phase diagrams of the optimal strategy but preserve their qualitative aspects. Finally, we also analyzed the least effective  strategies in a pool of four proposals and found out a  dependence on epidemiological parameters more complex than the case of the optimal strategy, in which the phase diagrams indicating the least effective strategy depends strongly on both IFR and demographics. {This is a result of the feedback loop between the disease dynamics and vaccination~\cite{Wang2016}, that must be taken into account by health authorities.}

The remaining of the paper is organized as follows. We describe the model and parameters used in the data-driven approach in section~\ref{sec:disease_model}. Results dealing with four prioritization strategies are shown in section~\ref{sec:results_and_discussion}. We conclude the paper discussing the main results in section~\ref{sec:conclusion}.

\section{Modeling the disease dynamics}
\label{sec:disease_model}

\subsection{SEAIRD compartmental model and contact matrices}
\label{subsec:model_vac}

\begin{figure*}[t]
	\centering 
	\includegraphics[width=0.75\linewidth]{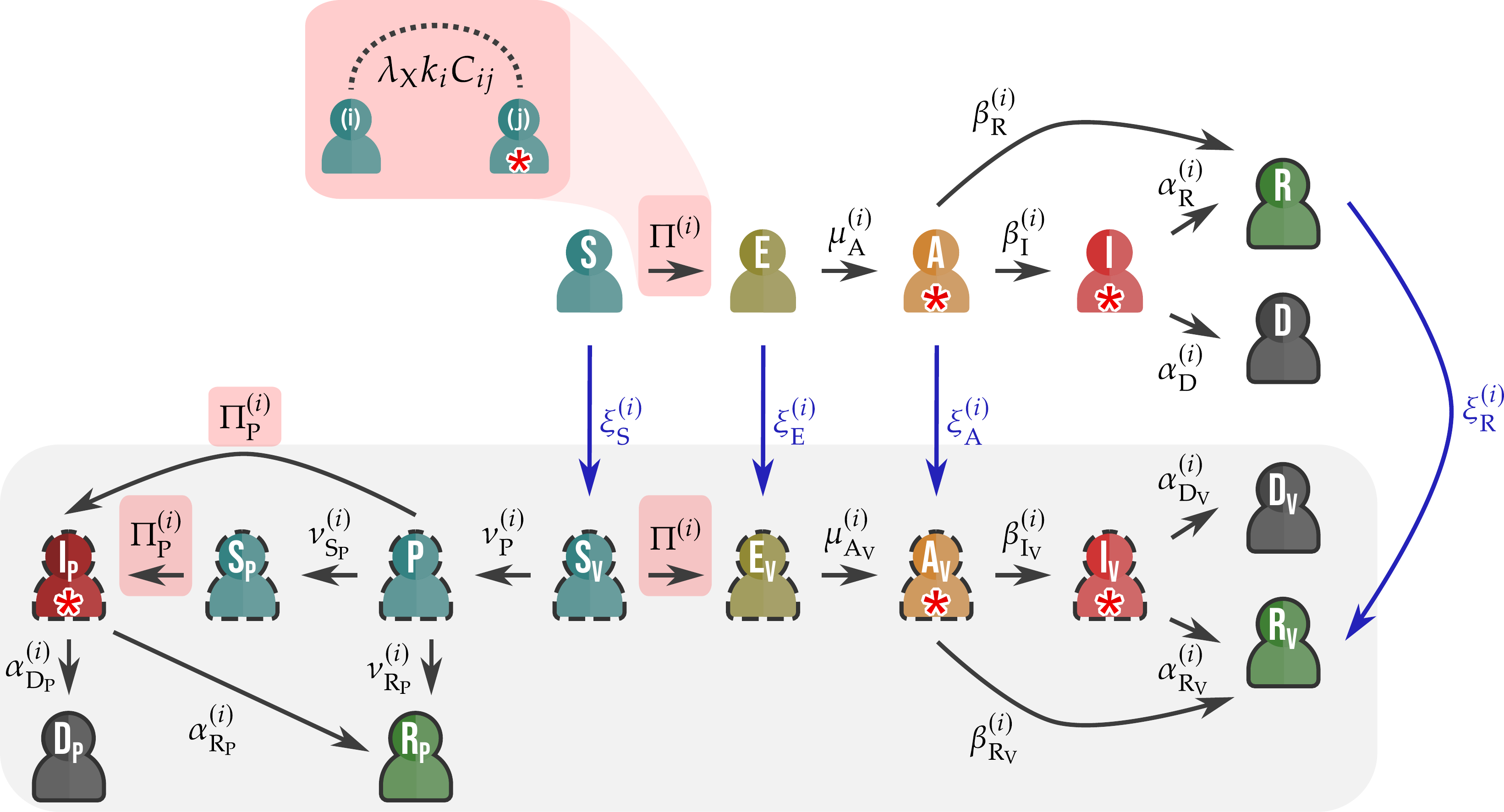}
	\caption{{Schematic representation of an age-structured compartmental model for epidemic {dynamics}  with vaccination.} The transitions between compartments {(labels are defined in the main text)} and the corresponding rates {(see Table~\ref{tab:params} and Eq.~\eqref{eq:Pi})} are indicated by arrows. The $\star$ symbol refers to the infectious individuals. The catalytic transitions {highlighted in the top} represent the contagions of susceptible individuals, vaccinated or not, upon infectious contacts, while the remaining transitions are spontaneous.} 
	\label{fig:model}
\end{figure*}

We start by considering a SEAIRD, age-structured, compartmental model~\cite{Rohani2007} to simulate the epidemic spreading without vaccination.  The schematic representation of this dynamics is shown in Fig.~\ref{fig:model}.  Individuals can belong to the following overall compartments: $\cS$ (susceptible), $\cE$  (exposed), $\cA$ (asymptomatic or presymptomatic), $\cI$ (infected), $\cR$ (recovered), and $\cD$ (deceased). Every compartment is subdivided according to the age group $i = 1,\ldots,~N_g$.  Susceptible individuals of age group $i$ become exposed with rate $\ci\Pi$ upon contact with infectious individuals (labeled by $\star$ in Fig.~\ref{fig:model}) of all age groups according to the contact matrix {to be defined below}. The other transitions are spontaneous: $\cE\rightarrow\cA$ happens with rate $\ci\mu_\cA$;  $\cA\rightarrow\cI$  and $\cA\rightarrow\cR$  have rates $\ci\beta_I$ and $\ci\beta_\cR$, respectively; those in the compartment $\cI$ recover from the disease or die with rates $\ci\alpha_\cR$ or $\ci\alpha_\cD$, respectively. Recovered individuals are assumed to be permanently immunized but still demand vaccines.

\begin{figure}[t!]
	\centering 
	\includegraphics[width=0.74\linewidth]{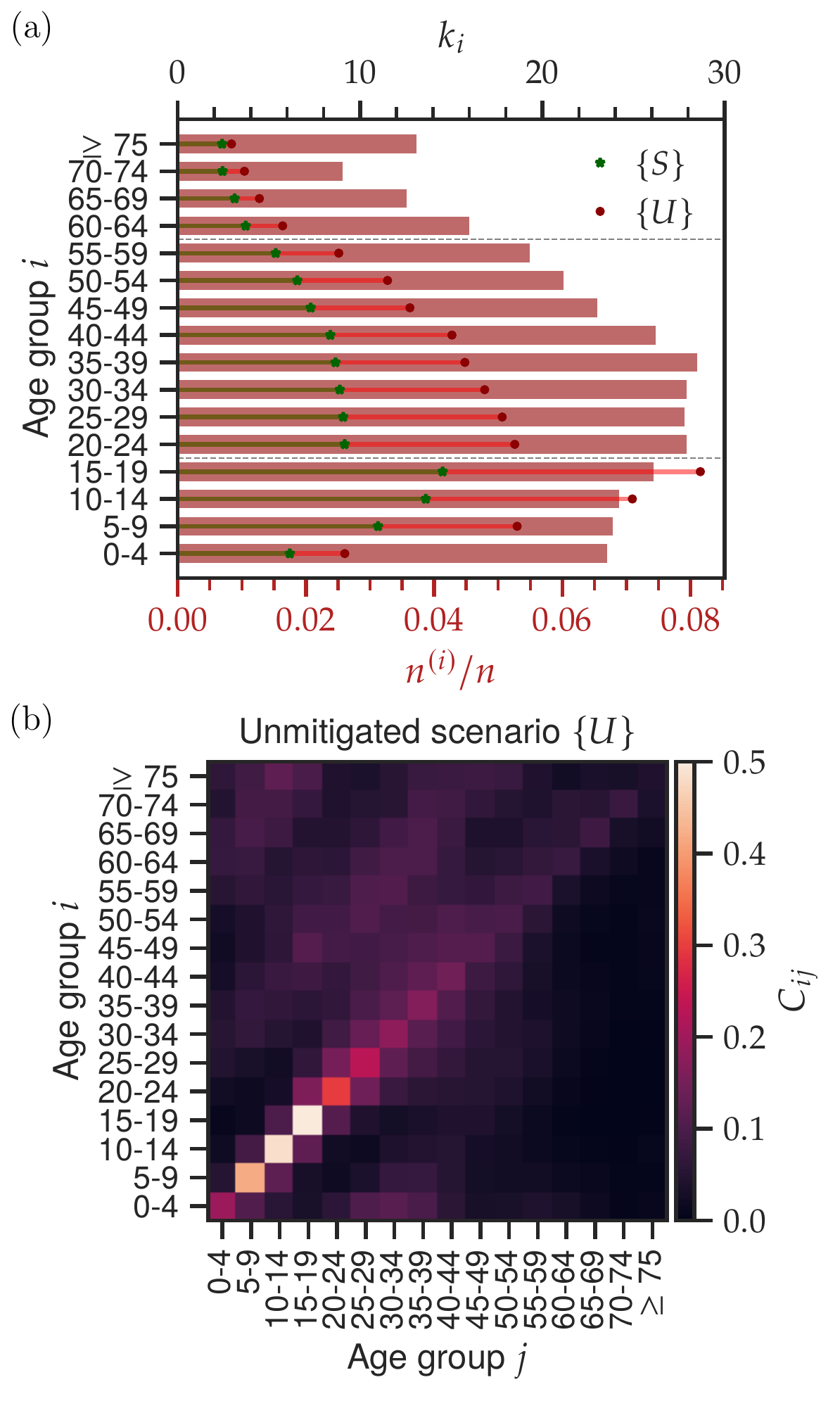}
	\caption{{Demographic and contact patterns in Brazil {considered in the age-structured compartmental model}.} (a) Fraction of individuals $\nni/n$ for each age group $i$ (bars) and their mean number of contacts $k_i$ (symbols) for social distancing $\uppX{S}$ and unmitigated $\uppX{U}$ scenarios. {Solid lines} represent the increase from one scenario to the other {and dashed lines divide the age groups into young, adult and elderly individuals}. (b) Contact matrix $C_{ij}$ in the unmitigated scenario of Brazil. Data adapted from Refs.~\cite{ibge2018projecoes,Prem2017} {as described in the main text}.}
	\label{fig:demographic}
\end{figure}

We consider individuals divided into $N_\text{g} = 16$ age groups, starting from 0--4, 5--9, up to 70--74 and $\geq 75$ years, {and  denote $\nni$ as the number of individuals in age group $i= 1,\ldots, 16$}. We refer to \textit{young} population as those with age 0 to 19, \textit{adults} for 20 to 59, and \textit{elderly} with age equal or above 60 years. The relative size of each group is given in Table~\ref{tab:demography} for three countries investigated in the present work. The estimates for the Brazilian demographics of 2020~\cite{ibge2018projecoes} are shown in Fig.~\ref{fig:demographic}(a).

\begin{table}[!h]
	\centering
	\caption{{ Distribution of the population of young, adult, and elderly individuals in Brazil~\cite{ibge2018projecoes}, Germany~\cite{united2019world}, and Uganda~\cite{united2019world}.}}
	\label{tab:demography}
	\setlength{\tabcolsep}{0.5em} 
	{\renewcommand{\arraystretch}{1.2}
		\begin{tabular}{cccc}
			\hline\hline
			& young & adult & elderly  \\
			\hline
			Brazil & 27.9\% & 57.6\% & 14.5\%  \\
			Germany & 18.9\% & 52.0\% & 29.1\%  \\
			Uganda & 57.1\% & 39.6\% & 3.3\%  \\
			\hline\hline
		\end{tabular}
	}
\end{table}

The infection rate of a susceptible individual within a group $i$ is given by
\begin{equation}
	\begin{split}
		\ci\Pi = \frac{k_i}{n} \sum_j { C_{ij}} \left( \lbcp{\cA} \cj{\icA} + \lbcp\cI \cj{\icI} +
				\right.\\
				\left.
		\lbcp\cAV \cj{\icAV} + \lbcp\cIV \cj{\icIV} + \lbcp\cIP\cj{\icIP}    \right),
	\end{split}
	\label{eq:Pi}
\end{equation}
where the corresponding {number of individuals in each compartment} $\cX\in\lbrace\cA,\cI,\cAV,\cIV,\cIP\rbrace$ and {age group $i$} is denoted by the italicized symbols {(\textit{e.g.} $\ci\icI$ is the number of infected  individuals belonging to age group $i$)}, while the corresponding infection rate per contact is represented by  $\lambda_\cX$. {In Eq.~\eqref{eq:Pi}, $k_i$ is the  number of contacts made by {an individual of} age group $i$, $C_{ij}$   is the  {fraction of contacts per individual of age} group $i$  {with those of group $j$,}  and $n$ is the total population.}

{To construct the contact matrices for $C_{ij}$ and $k_i$ we extracted the estimated number of contacts made by individuals of age groups $i$ and $j$, denoted by $m_{ij}$, from data  for three countries (Brazil, Uganda, and Germany) and 16 age groups (0--4, 5--9,$\cdots$, 75--79) reported in Ref.~\cite{Prem2017}. We assume that individuals with age $\geq 80$ follow the same contact patterns of those in group 75--79, considering a single age group $\geq 75$. Reference~\cite{Prem2017} reports matrices for contacts made at home, work, school, and other places in a situation without epidemic mitigation. The mitigation scenarios are modeled by the reduction of contacts in specific places. The values of $m_{ij}^\uppX{X}$ for a mitigation scenario $\uppX{X}$ are given by an weighted sum of the original matrices considering the fraction of contacts allowed in each place. It is important to ensure that the number of contacts among individuals of different groups are symmetric defining another contact matrix with values $c_{ij}$ where the total number of contacts made by individuals from age group $i$ with $j$, $\nni c_{ij}$, is the same as those of $j$ with $i$, $\nnj c_{ji}$. This is possible using~\cite{Medlock2009}}
\begin{equation}
	c_{ij} = \frac{m_{ij} \nni + m_{ji}\nnj}{2\nni}\,,
\end{equation}
%
%
{in which  $i,j = 1,\ldots, 16$ and obeys the balance condition $\nni c_{ij}= \nnj c_{ji}$.}
The average number of contacts made by individuals in age group $i$ is defined as
\begin{equation}
	k_i = \sum_{j} c_{ij}\,.
\end{equation}
Finally, the normalized contact matrix elements $C_{ij}$ are given by the relation $c_{ij} = k_i C_{ij}$ {and used in Eq.~\eqref{eq:Pi}}.

We  emulate a hypothetical epidemic scenario of \textit{social distancing}, denoted by $\uppX{S}$, where  100\%, 50\%, 50\%,  and 30\% of contacts are allowed in home, work, school, and other places, respectively. The social distancing will be compared with the unmitigated scenario, denoted by $\uppX{U}$, where no  reduction of contacts is implemented. The respective average number of contacts in different age groups for Brazil considering each scenario  is shown in Fig.~\ref{fig:demographic}(a). Adopting the symmetrization procedure  described before, we compute the contact matrices of Brazil shown in Fig.~\ref{fig:demographic}(b). For instance, the average number of contact decays from $\av{k} \approx 15$  in the unmitigated scenario to  $8.3$ when the social distancing is adopted. Equivalent figures for Uganda and Germany are provided in Section~SI-I of the {\SM}.

To simulate the epidemic dynamics we consider a set of ordinary differential equations considering  the transition rates and compartments schematically given in Fig.~\ref{fig:model}. 
The  evolution of the compartments of unvaccinated individuals is given by the following set of equations:
\begin{subequations}
	\label{eq:seaird}
	\begingroup
	\allowdisplaybreaks
	\begin{align}
		\dv{\ci{\icS}}{t} &= -  \ci\Pi \ci{\icS}   - \ri{\xi}{\cS}\ci\icS\,,\\
		\dv{\ci\icE}{t} &=  \ci\Pi \ci{\icS}  - \left( \ri\xi\cE +   \ri\mu\cA    \right) \ci\icE\,,\\
		\dv{\ci\icA}{t} &= \ri\mu\cA \ci\icE - \left( \ri\xi\cA + \ri\beta\cI + \ri\beta\cR  \right) \ci\icA\,,\\
		\dv{\ci\icI}{t} &= \ri\beta\cI \ci \icA - \left( \ri\alpha\cR + \ri\alpha\cD   \right)\ci\icI\,,\\
		\dv{\ci\icR}{t} &= \ri\beta\cR \ci\icA + \ri\alpha\cR \ci\icI - \ri\xi\cR \ci\icR\,,\\
		\dv{\ci\icD}{t} &= \ri\alpha\cD \ci\icI\,,
	\end{align}
	\endgroup
\end{subequations}
in which $\ci\Pi$ is given by Eq.~\eqref{eq:Pi} {and the italicized capital letters represent the number of individuals of age group $i$ in the compartment labeled with the same symbol}.

The basic reproduction number~\cite{Rohani2007} $R_0$  in absence of vaccines is given by the sum of the contributions from every infectious compartment, which are $\cA$ and $\cI$ individuals in the present work.
The contribution from asymptomatic infections is given by
\begin{equation}
	R_0^\cA =     \sum_{ij} \frac{\ci\icS}{n}
	a_j   k_i C_{ij} \frac{ \lbcp{\cA}     }{    \ri\beta\cI + \ri\beta\cR       }    \,,
\end{equation}
where $n=\sum_i\nni$ and $a_j$ is the probability that an infected individual is introduced in age group $j$ in a totally susceptible population, which is assumed to be proportional to the total number of contacts made by group $j$ and given by~\cite{Rohani2007}
\begin{equation}
	a_j = \frac{k_j \nnj}{\sum_l {k}_l \nnl} = \frac{{k}_j \nnj}{n \av{k}}\,.
\end{equation}
Here, $\av{k} = \sum_i k_i {\nni}/{n}$ is the average number of contacts. The infection rate $\lbcp{\cA}$ per contact is assumed to be the same for all age groups while  the mean time for which the individual of age group $i$ remains asymptomatic is given by $1/\left(\ri\beta\cI + \ri\beta\cR\right)$.  The final expression for $R_0^\cA$ becomes
\begin{equation}
	R_0^\cA =    \sum_{ij} \frac{  \lbcp{\cA} \nni \nnj k_i k_j C_{ij}    }{n^2 {\av{k}} \left(   \ri\beta\cI + \ri\beta\cR \right)}\,.
\end{equation}

For the infected compartment, we consider the probability that the introduced  individual becomes symptomatic, ${\ri\beta\cI}/{\left(\ri\beta\cI + \ri\beta\cR\right)}$,  and the mean time that {they remain} infectious, $1/\left(\ri\alpha\cD+ \ri\alpha\cR\right)$, and a similar calculation leads to
\begin{equation}
	R_0^\cI =   \sum_{ij}     \frac{\lbcp{\cI}}{  \ri\alpha\cD+ \ri\alpha\cR} \frac{\nni \nnj  k_i k_j C_{ij} \ri\beta\cI}{n^2 {\av{k}} \left( \ri\beta\cI + \ri\beta\cR \right)    }\,.
\end{equation}
Summing both contributions, we have:
\begin{equation}
	\label{eq:R0wosc_mt}
	R_0 = \sum_{i,j=1}^{N_\text{g}}  \frac{   \nni \nnj  k_i k_j C_{ij}      }{n^2 {\av{k}} \left(   \ri\beta\cI + \ri\beta\cR \right)} \left[ 
	\lbcp{\cA} + \frac{\lbcp{\cI}\ri\beta\cI}{  \ri\alpha\cD+ \ri\alpha\cR}
	\right]\,.
\end{equation}

\begin{figure}[h!]
	\centering 
	\includegraphics[width=0.99\linewidth]{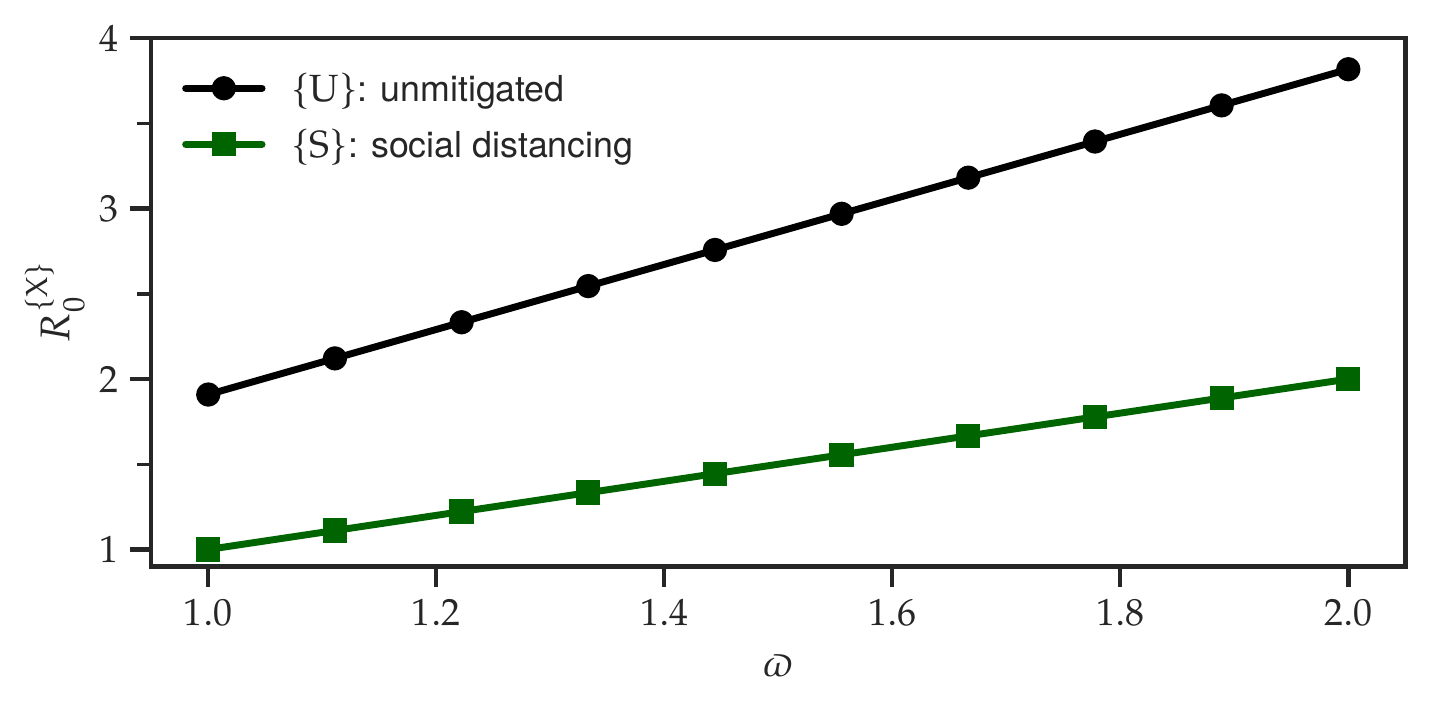}
	\caption{{Parameterization of the basic} reproduction number  $R_0^\uppX{X}$ {considering two different contact scenarios. The value of $R_0^\uppX{X}$ is computed using}  Eq.~\eqref{eq:R0wosc_mt} for {each scenario} $\uppX{X}$ (indicated in the legend) as function of the parameter $\lbpar$ for Brazil.}
	\label{fig:R0scenarios}
\end{figure}

Infection rate $\lambda$, assumed as $\lambda_\cX = \lambda$ for $\cX \in \lbrace\cA, \cI, \cAV,\cIV\rbrace$ and $\lambda_{\cIP} = \lambda/2$, is parameterized as a function of a control parameter $\lbpar$ using Eq.~\eqref{eq:R0wosc_mt} such that $\lbpar\equiv R_0^\uppX{S}$ for the case of social distancing scenario. {The parameter $\lbpar$ is  easier to interpret than $\lambda$ since it quantifies the level of the epidemic spreading in terms of a dimensionless quantity.} In the interval $\lbpar \in [1,2]$, the basic reproduction number of the unmitigated scenario is compatible with the SARS-CoV-2 ranges  $R_0\in [2,4]$ estimated at the {beginning} of the COVID-19 pandemics~\cite{Li2020,Zhang2020,Li2020a,Sanche2020}.
The computation of these values for both unmitigated and social distancing scenarios as a function of a parameter $\lbpar = R_0^\uppX{S}$, defined for a social distancing scenario, is shown in Fig.~\ref{fig:R0scenarios}. The interval $\lbpar \in [1,2]$ allows to mimic different levels of NPIs.

\subsection{Age-dependent infection fatality ratio}
\label{subsec:IFR}

\begin{figure}[h!]
	\centering 
	\includegraphics[width=0.99\linewidth]{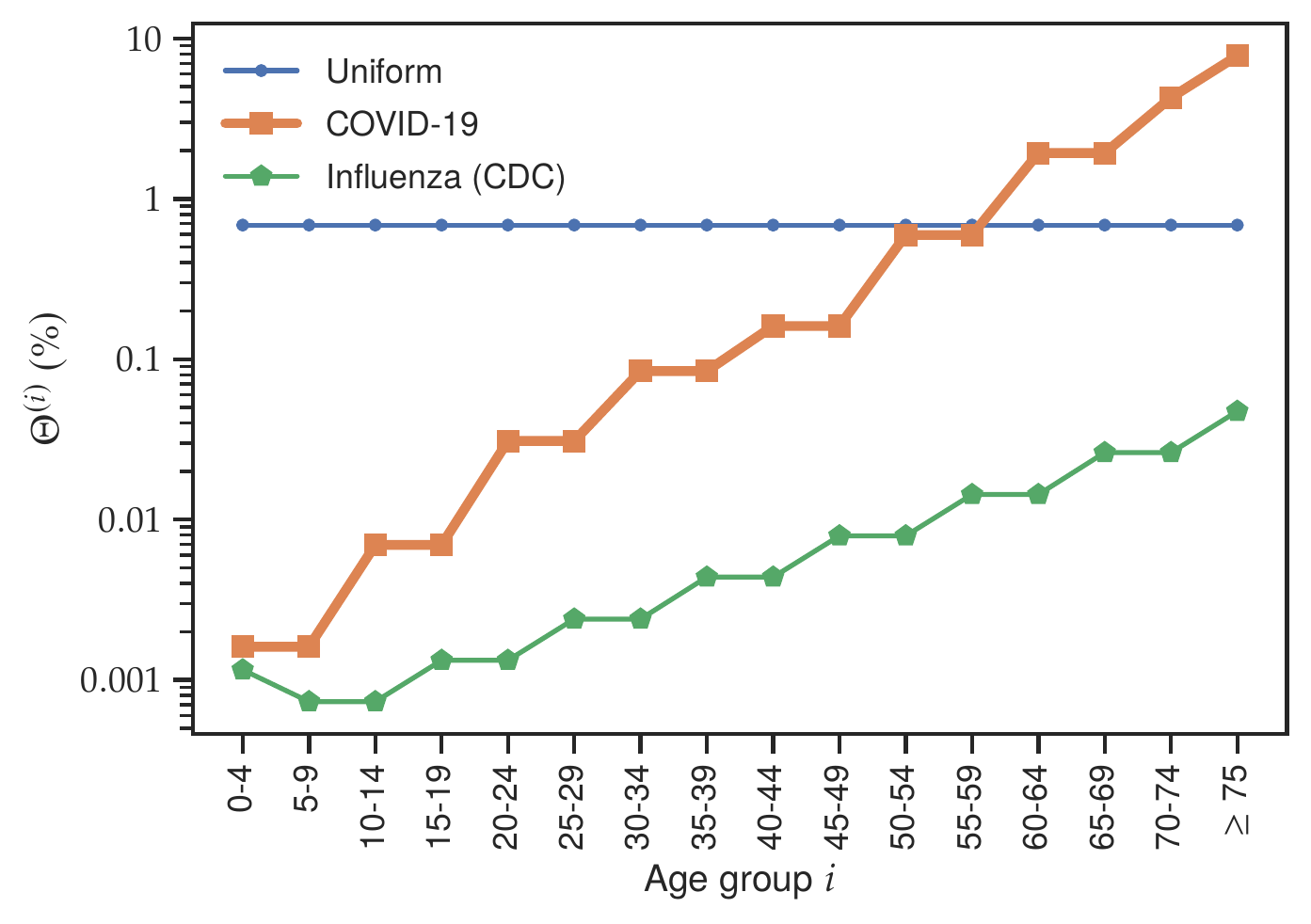}
	\caption{
	 {Age-dependent infection fatality ratio (IFR) for two different diseases and a baseline value. The IFR $\ci\Theta$ for each age group $i$ is shown for COVID-19~\cite{Verity2020,Poletti2020,Levin2020}, influenza~\cite{centers2020estimated}, and a uniform value of  0.68\% considered as a baseline, given by the average for COVID-19's IFR weighted by the population of each age group.}}
	\label{fig:fatality}
\end{figure}

The infection fatality ratio (IFR) considers all infections including the asymptomatic and paucisymptomatic ones, which may not be documented~\cite{Verity2020}, differently of the case fatality ratio, which is the fraction of confirmed cases that evolve to death. We define the IFR  as the fraction $\ci\Theta$ of infected individuals that evolve to death, and can be straightforwardly computed in terms of the model's rates (Fig.~\ref{fig:model}) as 
\begin{equation}
	\ci\Theta=\frac{\ci\beta_\cI}{\ci\beta_\cI+\ci\beta_\cR}\frac{\ci\alpha_\cD}{\ci\alpha_\cR+\ci\alpha_\cD}\,,
	\label{eq:theta_i}
\end{equation}
which can be {inverted} to determine the rate $\ci\alpha_\cD$ in terms of the IFR and other {experimentally determined} epidemiological parameters {and then $\ci\alpha_\cD$ is used in the simulation of the model equations}; See Table~\ref{tab:params}. Here we used the IFR estimates for COVID-19 reported by Verity \textit{et al}.~\cite{Verity2020}, that follows an exponential increase with age. Similar values have been reported elsewhere~\cite{Poletti2020,Levin2020}. To investigate the role of IFR, data for influenza from the \textit{Centers for Disease Control and Prevention}~\cite{centers2020estimated} and a hypothetical uniform IFR, given by the averaged COVID-19's IFR  weighted by the population size of each age group, were also considered while the remaining parameters were the same estimated for COVID-19. Influenza's IFR also increases exponentially with age, but it is lower than for COVID-19 and influences the choice of the vaccination strategies~\cite{Bubar2021,Fitzpatrick2021}. The IFR age dependence is shown in Fig.~\ref{fig:fatality}.

\subsection{Vaccination dynamics and strategies}
\label{sec:strategies}

Individuals of compartments $Z \in \{\cS, \cE, \cA, \cR\}$ can receive a vaccine shot and move to the corresponding vaccinated compartments $\cX_\cV \in \{ \cSV, \cEV, \cAV, \cRV\}$ with the respective rates $\ci\xi_{\cX}${; See Eq.~\eqref{eq:xiXinv}}. The transmission dynamics of vaccinated individuals belonging to $\cX_\cV$  are the same as $\cX$ with their respective transition rates. Additionally, those in the $\cSV$ compartment can turn into $\cP$ with rate $\ri{\nu}{\cP}$, conferring protection against {death} with high probability. The inverse of this rate is associated with the time between vaccine shots, an important epidemiological parameter.  The protected  individuals $\cP$ can evolve to either susceptible $\cSP$ or directly to recovered $\cRP$ states with rates $\ri{\nu}{\cSP}$ and $\ri{\nu}{\cRP}$, depending on the vaccine efficacy against infections $\ci\psi_\text{inf}$, respectively: $\cSP$ individuals can still be infected, transmit the pathogen, and eventually die, whereas $\cRP$ individuals are fully protected against both {infection and death}. The probability that a protected individual acquires full protection, given by 
\begin{equation}
	\ci\psi_\text{inf}=\frac{\ri{\nu}{\cRP}}{\ri{\nu}{\cRP}+\ri{\nu}{\cSP}}\,,
\end{equation}
is directly obtained from the model and used to parameterize the rates in terms of the vaccine efficacy against infection and a characteristic time $1/\left(\ri{\nu}{\cRP}+\ri{\nu}{\cSP}\right)$.
Both $\cP$ and $\cSP$ individuals can be infected with rate $\ci\Pi_\cP$ {which is explained below}. For the sake of simplicity, the infections of protected individuals follow a SIR-like dynamics with a single infectious compartment $\cIP$ which can be recovered  (move to $\cRP$) or die  (move to $\cDP$) with rates $\ri{\alpha}{\cRP}$ or $\ri{\alpha}{\cDP}$, respectively. The last two rates can also be parameterized in terms of the efficacy to prevent deaths which is the reduction of the probability that an infected protected individual dies in comparison with an unprotected population. {It can be calculated} in terms of the model rates {resulting in the following relation}
\begin{equation}
	\label{eq:deathprotectedrate}
	\left(1 - \ci\psi_\text{death}\right) \ci\Theta =  \frac{\ri{\alpha}{\cDP}}{\ri{\alpha}{\cRP}+\ri{\alpha}{\cDP}}\,,
\end{equation}
in which $\ci\Theta$ is the IFR of group age $i$. {Using the experimentally determined epidemiological parameters, IFR, and efficacy against death we then determine $\ri{\alpha}{\cDP}$ from  Eq.~\eqref{eq:deathprotectedrate}; See Table~\ref{tab:params}.}

\begin{figure}[t]
	\centering 
	\includegraphics[width=0.9\linewidth]{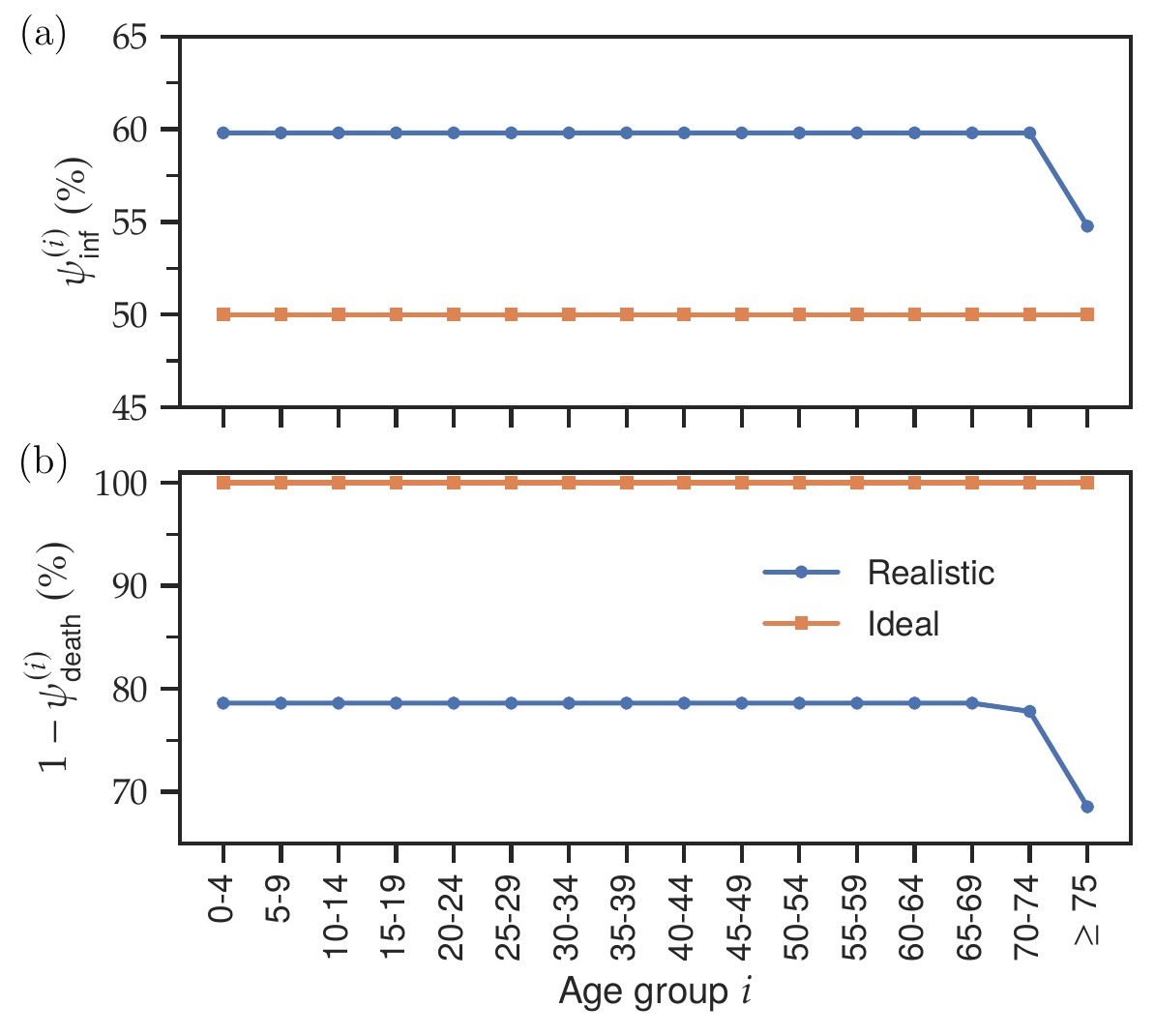}
	\caption{{Age-dependent vaccine efficacy against infection and death for realistic and ideal vaccines. The values of protection (a) against infection, $\psi_\text{inf}$, and (b) against death, $1-\psi_\text{death}$, are given for the realistic and ideal vaccine models. The realistic values are based in Ref.~\cite{CerqueiraSilva2021} and details are given in the main text.}}
	\label{fig:protection}
\end{figure}

For sake of simplicity, individuals are labeled as vaccinated after the first shot, but acquire protection and are moved to the compartment $\ci\cP$ only after the second shot. The average interval between shots, which corresponds to the transition $\cSV\rightarrow\cP$ is given by  $1/\ri{\nu}{\cSP}$, assumed uniform across age groups. With respect to the efficacy against infection and death, both an ideal case with moderate protection ($\ci{\psi}_\text{inf} = 50$\%$~\forall~i$) against infection with full protection against death ($\ri\alpha\cDP = 0~\forall~i$), and an age-dependent efficacy based in the values reported in Ref.~\cite{CerqueiraSilva2021}, for which elderly individuals have reduced protection, are simulated. 
{We considered data for CoronaVac effectiveness against infection and death for individuals who received two shots, based on values reported in Ref.~\cite{CerqueiraSilva2021}, which are available for age groups  $<60$, 60--69, 70--79, 80--89 and $\geq 90$ years old. We assume a constant efficacy against infection for individuals with $<75$ years, and constant efficacy against death for $<70$ years. 
The reported value for 70--79 is used as the efficacy against death for the age group 70--74, while an average weighted by the populations is used to obtain the values for individuals of $\geq 75$ years in both cases.}
Realistic and ideal models for protection are shown in Fig.~\ref{fig:protection}. In all cases, after an average time of $1/\left({\nu}_{\cRP}+{\nu}_{\cSP}\right)=7$~days, uniform across ages, they  either become fully immunized ($\cP\rightarrow \cRP$) or remain susceptible to the disease ($\cP\rightarrow\cSP$). 

\begin{table*}[t!]
	\caption{Epidemiological parameters and  rates used in the model.}
	\label{tab:params}
	\renewcommand{\arraystretch}{1.2}
	\small
		\begin{tabular}{
				p{.15\textwidth-2\tabcolsep}
				p{.5\textwidth-2\tabcolsep}
				p{.23\textwidth-2\tabcolsep}
				p{.13\textwidth-2\tabcolsep}
			}
			Parameters  & Description & Value & References \\\hline
			$\lambda_\cX$, --- & Transmission rates for $\cX = \cA, \cI, \cAV,$ and $\cIV$  &  See Eq.~\eqref{eq:R0wosc_mt} \vspace{0.5em}&    ---   \\
			---, $\lambda_{\cIP}$ &Transmission rate for partially protected individuals &  $\ds\frac{\lambda_\cX}{2}$ \vspace{0.5em} & \cite{LevineTiefenbrun2021} \\
			$\ri{\mu}{\cA}$, $\ri{\mu}{\cAV}$ &  Latent period rate &      $\left(5.2~\text{days}\right)^{-1}$     &  \cite{Li2020}     \\
			$\ri{\beta}{\cI}$,  $\ri{\beta}{\cIV}$ & Asymptomatic period rate &  $\left(2.6~\text{days}\right)^{-1}$        &    \cite{Li2020}  \\
			$\ri{\alpha}{\cR}$, $\ri{\alpha}{\cRV}$ &   Recovering rate (with symptoms)  &      $\left(3.2~\text{days}\right)^{-1}$    \vspace{0.5em}   &   \cite{Read2021,Danon2021}    \\
			$\ri{\beta}{\cR}$, $\ri{\beta}{\cRV}$ &  Recovering rate (without symptoms) &     
			$\ds\frac{1}{{{\beta}_{\cI}}^{-1} + {{\alpha}_{\cR}}^{-1} }$
			&    ---    \\
			$\ri{\alpha}{\cD}$,  	$\ri{\alpha}{\cDV}$ & Death rate  &    See Eq.~\eqref{eq:theta_i}      \vspace{0.5em}           &    ---   \\
			---, $\ri{\nu}{\cP}$ &  Immune response period rate  &  $\left(21~\text{days}\right)^{-1}$ \vspace{0.5em}   & \cite{Chodick2021,Maier2021} \\
			---, $\ri{\nu}{\cRP}$ &  Vaccine success rate  &  $\ds\frac{1}{7~\text{days}}\ci{\psi}_\text{inf} $\vspace{0.5em}  & --- \\
			---, $\ri{\nu}{\cSP}$  &   Vaccine failing rate   &     $\ds\frac{1}{7~\text{days}}\left( 1 - \ci{\psi}_\text{inf} \right)$       & --- \\
			---, $\ri{\alpha}{\cRP}$ &  Recovering rate for protected individuals & 
			$\ri{\beta}{\cR}$ & ---
			\\
			---, $\ri{\alpha}{\cDP}$ &  Death rate for protected individuals &    See Eq.~\eqref{eq:deathprotectedrate} & ---    \\
			$\ri\xi\cX$, --- & Vaccination rate for a given compartment $\cX$  & See Eq.~\eqref{eq:xiXinv} & --- \\\hline
			$\ci{\Theta}$ & Infection fatality ratio & See Fig.~\ref{fig:fatality} & \cite{Verity2020,centers2020estimated} \\
			$\ci{\psi}_\text{inf}$ & Vaccine efficacy against infection &  See Fig.~\ref{fig:protection}  & \cite{CerqueiraSilva2021} \\
			$\ci{\psi}_\text{death}$ & Vaccine efficacy against death & See  Fig.~\ref{fig:protection}  & \cite{CerqueiraSilva2021} 
		\end{tabular}
\end{table*}

The dynamics of vaccinated individuals is highlighted in the shaded area of Fig.~\ref{fig:model} and the temporal evolution of the vaccinated compartments is given by the following set of equations:
\begin{subequations}
	\label{eq:seairdV}
	\begingroup
	\allowdisplaybreaks
	\begin{align}
		\dv{\ci{\icSV}}{t} &=  -  \left( \ci\Pi  + \ri\nu\cP \right) \ci{\icSV}  + \ri\xi\cS \ci\icS  \,,\\
		\dv{\ci{\icEV}}{t} &= \ci\Pi  \ci{\icSV} + \ri\xi\cE \ci\icE - \ri\mu\cAV \ci\icEV \,,\\
		\dv{\ci{\icAV}}{t} &=\ri\mu\cAV \ci\icEV + \ri\xi\cA \ci\icA - \left( \ri\beta\cIV + \ri\beta\cRV  \right) \ci\icAV \,,\\
		\dv{\ci{\icIV}}{t} &= \ri\beta\cIV \ci\icAV  - \left(  \ri\alpha\cRV + \ri\alpha\cDV   \right) \ci\icIV\,,\\
		\dv{\ci{\icRV}}{t} &= \ri\alpha\cRV\ci\icIV + \ri\beta\cRV\ci\icAV + \ri\xi\cR \ci\icR \,,\\
		\dv{\ci{\icDV}}{t} &= \ri\alpha\cDV \ci\icIV\,.
	\end{align}
	\endgroup 
\end{subequations}
Finally, individuals that develop immune response to the disease are protected and can turn into  $\cP$, $\cSP$, $\cIP$, $\cRP$, and $\cDP$ compartments, which evolve as
\begin{subequations}
	\label{eq:xV}
	\begingroup
	\allowdisplaybreaks
	\begin{align}
		\dv{\ci{\icP}}{t} &=  \ri\nu\cP \ci\icSV - \left(   \ri\Pi\cP  + \ri\nu\cSP + \ri\nu\cRP   \right) \ci\icP \,,\\
		\dv{\ci{\icSP}}{t} &= \ri\nu\cSP \ci\icP - \ri\Pi\cP \ci\icSP \,,\\
		\dv{\ci{\icIP}}{t} &= \ri\Pi\cP \ci\icP + \ri\Pi\cP \ci\icSP- \left(\ri\alpha\cRP + \ri\alpha\cDP  \right)\ci\icIP\,,\\
		\dv{\ci{\icRP}}{t} &= \ri\nu\cRP \ci\icP + \ri\alpha\cRP \ci\icIP\,,\\
		\dv{\ci{\icDP}}{t} &= \ri\alpha\cDP \ci\icIP.
	\end{align}
	\endgroup
\end{subequations}
{For sake of simplicity, we consider the same infection rate for protected and non-vaccinated susceptible individuals, $\ri\Pi\cP = \ci\Pi$, assuming the former is  less contagions ($\lambda_{\cIP}<\lambda_{\cX}$) due to a reduced viral load~\cite{LevineTiefenbrun2021}.}

The rate $\xi(t)$ is defined as the per capita number of daily first shots of vaccines. For sake of generality, it is assumed to be time-dependent even though we performed simulations assuming $\xi$ constant in the present work. Let us define $\ci\Omega(t)=1$ if age group $i$ is being vaccinated at time $t$ and  $\ci\Omega(t)=0$ otherwise.  All non-vaccinated individuals belonging to the compartments  $\ci\cS, \ci\cE, \ci\cA$, or $\ci\cR$ can receive their first shots with equal chance if they are scheduled, \textit{i.e.}, if $\ci\Omega(t)=1$. Therefore, the vaccination rates of the compartments  $\cX \in\{ \cS, \cE, \cA,\cR\}$ in  age group $i$ are given by
\begin{equation}
	\ri\xi\cX(t) = 
		\frac{n\xi  { \ci\Omega}}{\sum_j \left[ \cj{S} + \cj{E} + \cj{A}  + \cj{R} \right]\cj\Omega}\,.
		\label{eq:xiXinv}
	\end{equation}
{For example, $\ri\xi\cS(t)$ is the vaccination rate of susceptible individuals of age group $i$.}
	
The prioritization of vaccine shots across different age groups  over time can be modeled  with $\ci\Omega(t)$. The vaccination begins at a time $t_\text{v}$ such that $\ci\Omega(t<t_\text{v})=0~\forall~i$. Once $80\%$ of a priority group has been vaccinated, the vaccination of the next priority groups starts concomitantly with all other groups where vaccination had started previously. 
Four prioritization strategies are investigated in the present work. In \textit{decreasing age priority} (DAP)  strategy, one starts in the oldest age group and proceeds progressively down to the youngest one as  adopted in many countries for the general population. In \textit{highly-vulnerable priority} (HVP) strategy,  only the elderly are prioritized according to the age, then all adults (age 20--59) and later all young individuals (0--19) are vaccinated  without age prioritization. This could represent the economically active population being vaccinated altogether after the most vulnerable individuals were protected.  The \textit{decreasing contact priority} (DCP) strategy starts with the age group of the higher number of contacts and proceeds progressively down to the one less connected. This strategy corresponds to vaccinating the most exposed first. Finally, in the \textit{no priority} (NP) strategy all age groups are vaccinated concomitantly.

	
Finally, we integrate the dynamical system  using  initial conditions where a single exposed individual is introduced in a single age group $s$ in a total population $n=10^5$ individuals (results are insensitive to this parameter given it is large enough).  The averages were computed over initial conditions $s=1,\ldots,N_\text{g}$ using as weight the total number of contacts made by each group $s$, in which the epidemic process is initiated. 	Table~\ref{tab:params} presents all other epidemiological parameters used in the model or their respective relations.

\section{Results and discussion}
\label{sec:results_and_discussion}
	
\subsection{DAP vaccination}
\label{sec:dap_vac}
	
\begin{figure*}[t]
	\centering 
	\includegraphics[width=0.75\linewidth]{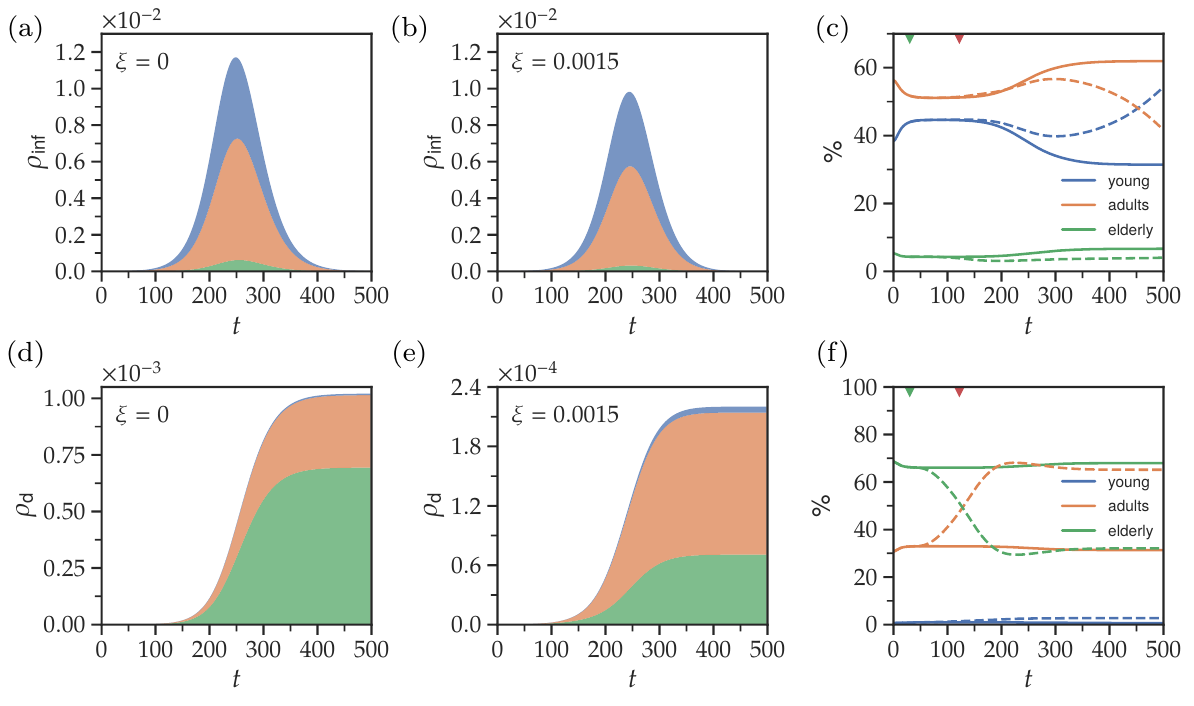}
	\caption{{{Changes in the age} profiles of infectious individuals and deaths with and without vaccination.} Evolution of the fraction and age profiles of (a-c) infectious individuals and (d-e) accumulated deaths. The effects of DAP  strategy with age-dependent values of efficacy against infection and death are addressed with a fixed vaccination rate $\xi = 0.15\%$ per day and delay of $t_\text{v} = 30$~days in a scenario of social distancing  with $\lbpar = R_0^\uppX{S}= 1.3$. In the stack plots, the envelope gives the total prevalence or deaths while colors give the proportion within the young (0--19 yr, blue), adult (20--59 yr, orange), and elderly ($\geq 60$ yr, green) age groups.  The age profiles for (c) infectious and (f) accumulated deaths give the percentage distribution for each group with (dashed lines) and without (solid lines) vaccination. Triangles indicate when the vaccinations of the elderly and adult populations start, while {for} the young population it has not started in the investigated time window. Note that the scales in (d) and (e) are different.}
		\label{fig:ageprofile}
	\end{figure*}

Considering $\lbpar = 1.3$, social distancing scenario,  modest  vaccination with constant rate $\xi=0.15\%$  of the population per day, age-dependent  efficacy, short delay $t_\mathrm{v} = 30$ days, and DAP vaccination strategy,  we computed the fractions $\rhoinf$ of  infectious ($\cI$, $\cIV$, $\cA$, $\cAV$, and $\cIP$)  and  $\rhod$ of deceased ($\cD$, $\cDV$, and $\cDP$) individuals to compare with the case without vaccines. Figures~\ref{fig:ageprofile}(a,b,d,e) present the temporal evolution of $\rhoinf$ and $\rhod$ in the form of stack plots split  according to the age profiles of young, adult, and elderly populations of Brazil. Even a modest vaccination rate,  if started early, substantially reduces the total  amount of deaths while the reduction of infections is not expressive. The age profiles for infections and deaths are shown in Figs.~\ref{fig:ageprofile}(c,f). Since the percentages of young, adults, and elderly in the Brazilian populations  are 27.9\%, 57.6\%, and 14.5\%~\cite{ibge2018projecoes}, respectively, the age profile of infections without vaccines is  highly correlated with  the demography, Fig.~\ref{fig:ageprofile}(c), while the deaths' profile is determined by the COVID-19's IFR used in these simulations: Figs.~\ref{fig:fatality} and \ref{fig:ageprofile}(f) show that deaths are highly concentrated in elderly, even they corresponding to the minor part of the population.

The DAP strategy moderately alters the age profile of infected individuals. Beyond reducing deaths in all age groups, the fatality age profile is highly affected, presenting a big drop in death among the elderly and  a fractional rise in the adult population that now concentrates most of the deaths. Not surprisingly, this inversion was observed during the first semester of 2021 in Brazil that adopted DAP after vaccination of healthcare workers and persons with morbid conditions. {Remark that the age profile of deaths changes substantially after the lifespan of immune response $1/\ri{\nu}{\cP} = (21~\mathrm{days})^{-1}$, highlighting the importance of a complete immunization scheme.} Similar results are found using uniform values of efficacy against death and infection, as shown in Section~SI-II of the \SM, with quantitative changes on the age profile due to the reduction of the drop of efficacy in the elderly individuals.
	
\begin{figure}[b!]
\centering 
\includegraphics[width=0.99\linewidth]{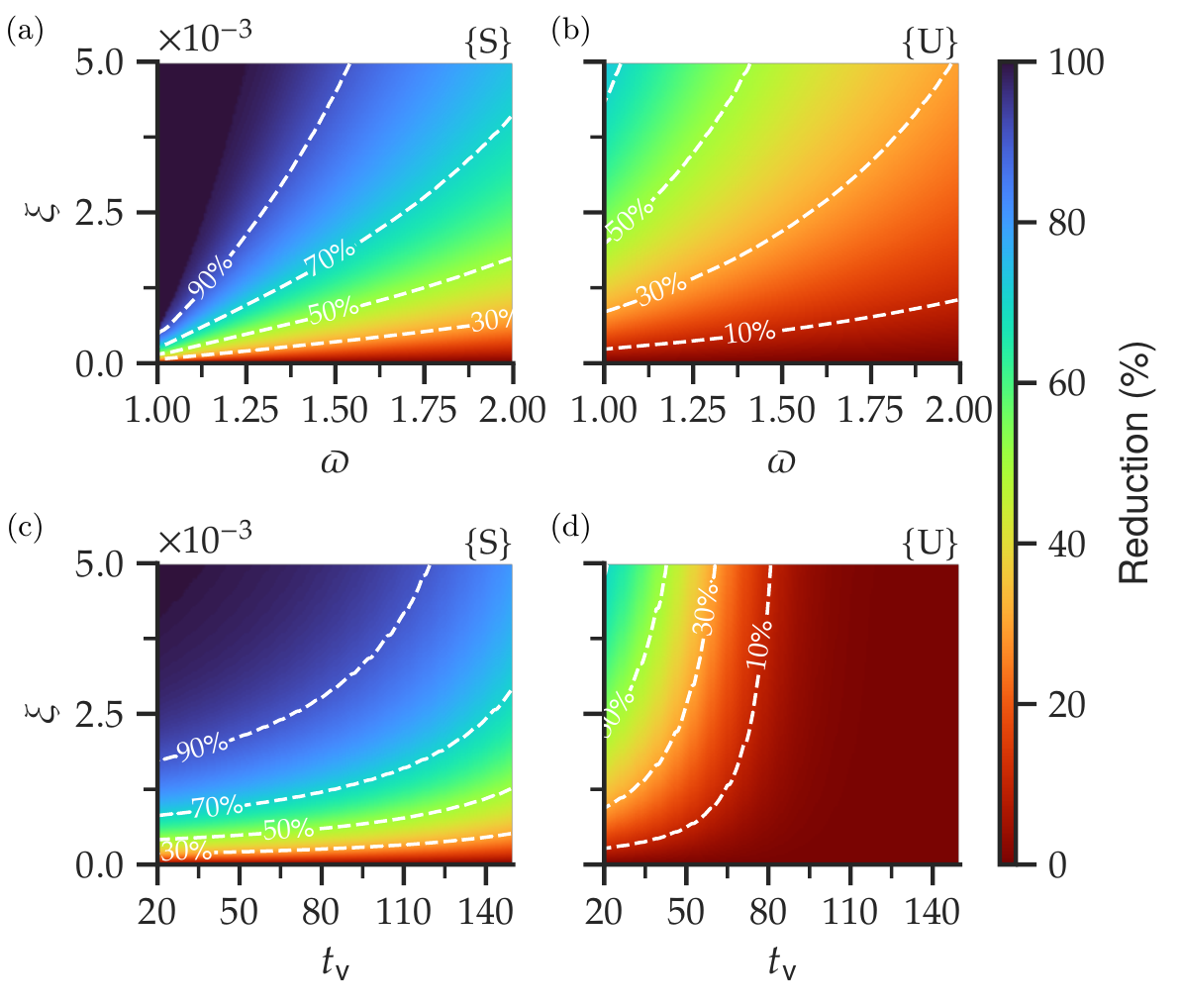}
\caption{{Effects of infection rate and time delays for different vaccination rates in reducing the number of deaths. Heatmaps and isolines (white curves)  for reduction of deaths using a DAP strategy and age-dependent values of efficacy against death and infection in comparison with the situation without vaccines.}  Fixed delay of  $t_\text{v} = 30$ days is considered in (a,b) while a fixed infection rate given by $\lbpar = 1.3$ is considered in (c,d).  Social distancing $\uppX{S}$ and unmitigated  $\uppX{U}$ scenarios are presented in left and right-hand panels, respectively.}
\label{fig:reduction_scenarios}
\end{figure}
	
The interplay between vaccination rate $\xi$ and effective infection rate parameterized by $\lbpar$ is investigated considering the DAP strategy and age-dependent values of efficacy with a delay of $t_\text{v}= 30$~days to start vaccination. Heatmaps for the reduction of deaths in the space parameter $\xi$ versus $\lbpar$ under unmitigated and social distance contact scenarios are presented in Fig.~\ref{fig:reduction_scenarios} for Brazil contact matrices. The corresponding heatmaps for total recovered population is given in Fig.~SI-4 of the {\SM}. As expected, the reduction of deaths is much more expressive than of infections for both scenarios. In the  scenario of social distancing shown in Fig.~\ref{fig:reduction_scenarios}(a), DAP vaccination performs very well to reduce deaths if the immunization rate is not too low and dissemination rate is not too high (reduced $\lbpar$); the latter is feasible through simple NPIs. For the case of unmitigated contacts shown in Fig.~\ref{fig:reduction_scenarios}(b), the vaccination can significantly reduce the number of deaths only at a high vaccination rate of $\xi=0.5$\% per day (approximately seven months for the total population to be immunized), only if infection rate is kept near to the lower bound of $\lbpar$. The unmitigated scenario is not able to reduce the transmission  by more than 10\% in the whole parameter space, while we can still see a significant effect in the social distancing scenario; See Fig.~SI-4 of the \SM. The previous discussion was done for an early beginning of the vaccination. However, the intervention time $t_\text{v}$ is a key parameter to the effectiveness of the vaccination. Heatmaps of death reduction in the $\xi$ versus $t_\text{v}$ space's parameter are presented in Fig.~\ref{fig:reduction_scenarios}(c) and (d) for $\lbpar = 1.3$ ($R_0>1$ for $\xi = 0$ in both investigated scenarios). Delays are extremely harmful to the vaccination effectiveness  even in the social distancing scenario shown in Fig.~\ref{fig:reduction_scenarios}(c), in which one sees that the death reduction drops {significantly} for  $t_\text{v}\gtrsim 80$~days, the more for lower vaccination rate. Delayed vaccination becomes ineffective in the scenarios without mitigation even at a high vaccination rate, as shown in Fig.~\ref{fig:reduction_scenarios}(d).
	
\subsection{Comparing strategies}
\label{sec:comp_st}

\begin{figure}[h!]
\centering 
\includegraphics[width=0.99\linewidth]{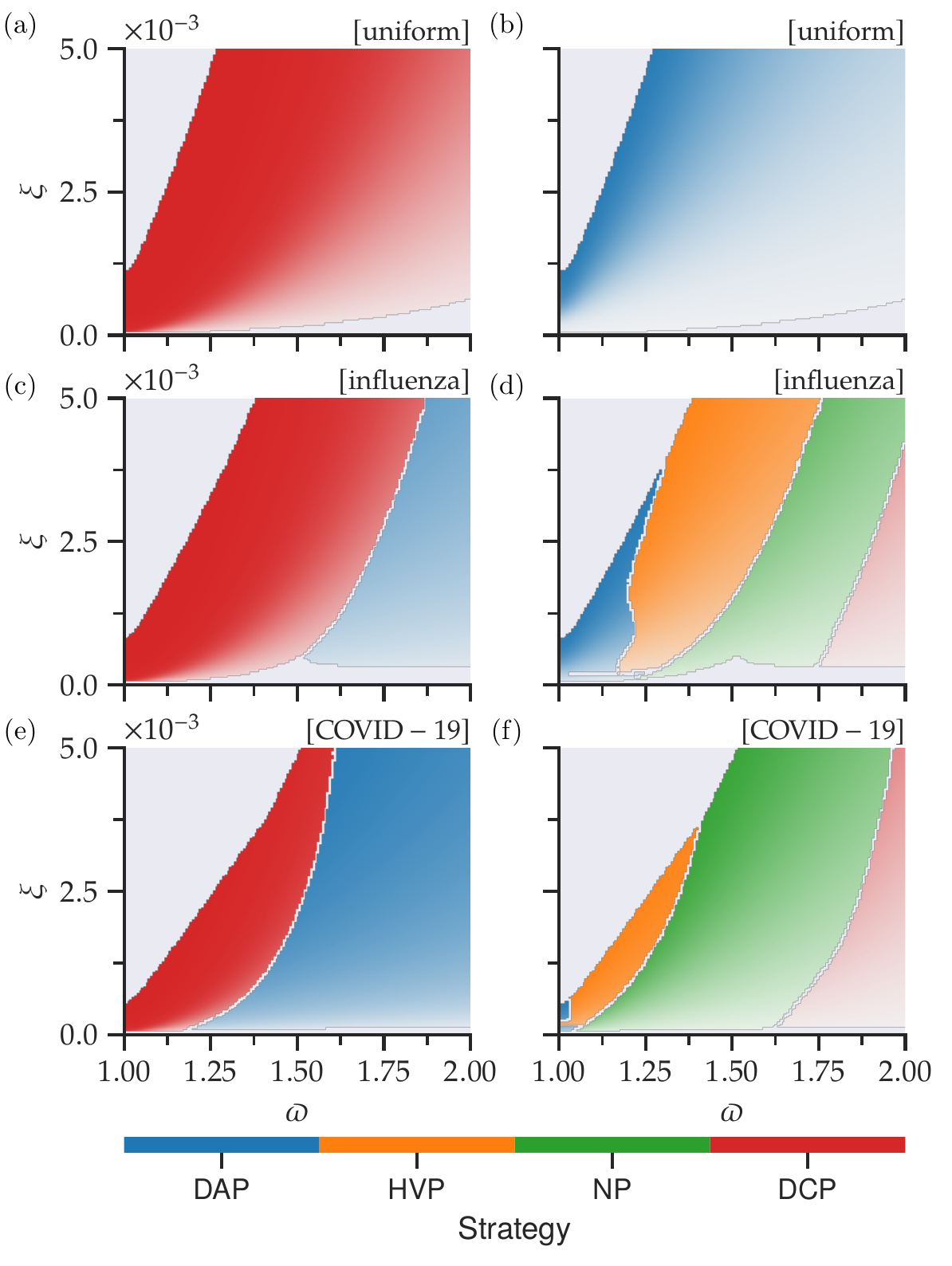}
\caption{{Comparison of the most and least effective strategies for different IFR age profiles}. The diagrams indicate the most (left) and least (right) effective strategies for reduction  of deaths in the total population in the space parameter $\xi$ versus $\lbpar$.  Three IFR age profiles are considered: (a,b) uniform, (c,d) influenza, and (e,f) COVID-19; See Fig.~\ref{fig:fatality}. Four vaccination strategies  are considered: decreasing age (DAP), highly vulnerable (HVP), no (NP), and decreasing contact (DCP) priorities.  A time delay of $t_\mathrm{v}= 30$~days, vaccination with uniform values of efficacy against death and infection, Brazilian demography, and social distancing contact scenario  were considered. The gradient colors refer to the respective reduction of deaths, the darker the higher. Differences between the most and least effective strategies smaller than 5\% are depicted in gray.}
\label{fig:best_with_all}
\end{figure}
	
Given the limited availability of vaccine shots, an essential problem is to determine which prioritization strategy can be more effective. {Aiming at saving the maximum number of lives, we compared different strategies in the parameter space $\xi\times \lbpar$ shown in Fig.~\ref{fig:best_with_all} for Brazilian demography (Fig.~\ref{fig:demographic}) and uniform values of vaccine efficacy against death and infection. To isolate the effects of different IFR age profiles (Fig.~\ref{fig:fatality}), the same epidemiological parameters of COVID-19, except the IFR itself, were considered.} For the unreal case of age-independent IFR, the optimal strategy is to prioritize those who are more exposed, \textit{i.e.} make more contacts, using DCP followed by no priority; the latter is better than the remaining ones since the adults constitute simultaneously the largest and most connected populations in Brazil. For influenza's IFR, we observe that prioritizing the most exposed individuals is more advantageous than the most vulnerable population for a broader region of the space parameter, especially if the vaccination rate is high. However, for uncontrolled transmission (high $\lbpar$), it is still more effective to vaccinate according to a decreasing age criterion. Finally, the simulations with COVID-19's IFR yield that the most effective strategy is DAP in most of the investigated parameter space. Only in a narrow region, prioritizing the most exposed through DCP is the most effective. Notice that prioritizing only the highly vulnerable (elderly) individuals by adopting the HVP strategy is not the most effective strategy in the investigated diagrams. 
		
We also compute the least effective among the four investigated strategies, as shown in the right column of Fig.~\ref{fig:best_with_all}. For the uniform IFR age profile, the DAP strategy  reduces deaths least while for realistic IFR of influenza and COVID-19, a complex pattern emerges in the diagrams. HVP  can perform worst if the infection is moderately uncontrolled while DAP is the least effective only for almost controlled spreading ($\lbpar=R_0^\uppX{S}\approx 1$).  A remarkable result is that prioritization of the most exposed  is the least effective strategy if the epidemic is out of control. The respective plots for Uganda and Germany are shown in Fig.~SI-5  of the \SM.
	
\begin{figure}[h!]
\centering 
\includegraphics[width=0.99\linewidth]{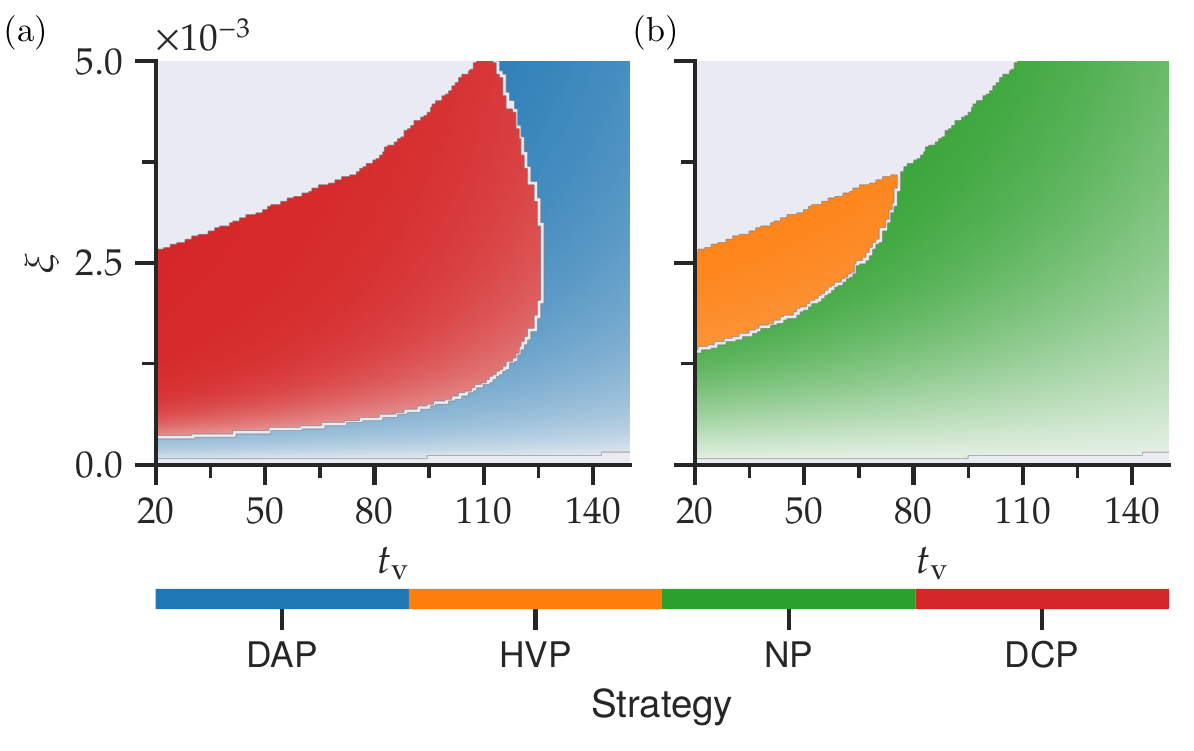}
\caption{{Comparison of the most and least effective strategies considering different time delays.} Diagrams indicating the (a) most  and (b) least effective  strategies to reduce deaths in  the parameter space  $\xi$ versus $t_\mathrm{v}$ considering four strategies. The COVID-19' IFR age profile, contact scenario of social distancing, vaccination with age-dependent values of efficacy against death and infection, Brazilian demography, and infection rate given by  and $\lbpar = 1.3$ were considered.  Colors as in Fig.~\ref{fig:best_with_all}.}
\label{fig:best_worst_tv}
\end{figure}

The role of time delay in the strategy effectiveness is presented  in Fig.~\ref{fig:best_worst_tv} with fixed infection parameter $\lbpar = R_0^\uppX{S} = 1.3$ in a social distancing scenario, COVID-19's IFR age profile, and age-dependent values of efficiency. As shown in Fig.~\ref{fig:best_worst_tv}(a), DAP is the most effective for large delays while DCP is for earlier interventions. Moreover, the re-entrant behavior for intermediate delays ($t_\text{v}\sim 100$ days) reveals a complex interplay between epidemiological parameters and optimal strategies. No prioritization has the worst performance in almost the entire parameter diagram; See Fig. \ref{fig:best_worst_tv}(b). Despite the unavoidable ethics concerns  in prioritization strategies, a remarkable feature of these diagrams is that the optimal strategy depends, in a very  nonlinear fashion, on the level of epidemic transmission, immunization rates, and timeliness of starting the vaccination.

\subsection{Effect of social contacts and demography}
\label{sec:diff_countries}
	
We have so far observed the nonlinear interplay between epidemiological parameters to determine the optimal strategy to reduce deaths. It depends nontrivially on infection scenarios, vaccination rates, efficacy, and IFR age profiles. Now we explore the effect of demography and contact structures considering two other countries: Uganda, with a higher number of young individuals, and Germany, with a higher number of elderly individuals in comparison with Brazil. Table~\ref{tab:demography} summarizes the differences between young, adult and elderly  populations in these countries while detailed demographics and contact patterns for Uganda and Germany are shown in Figs.~SI-1 and SI-2 of the {\SM}  in complement to Fig.~\ref{fig:demographic} for Brazil.
	
\begin{figure}[t!]
\centering 
\includegraphics[width=0.99\linewidth]{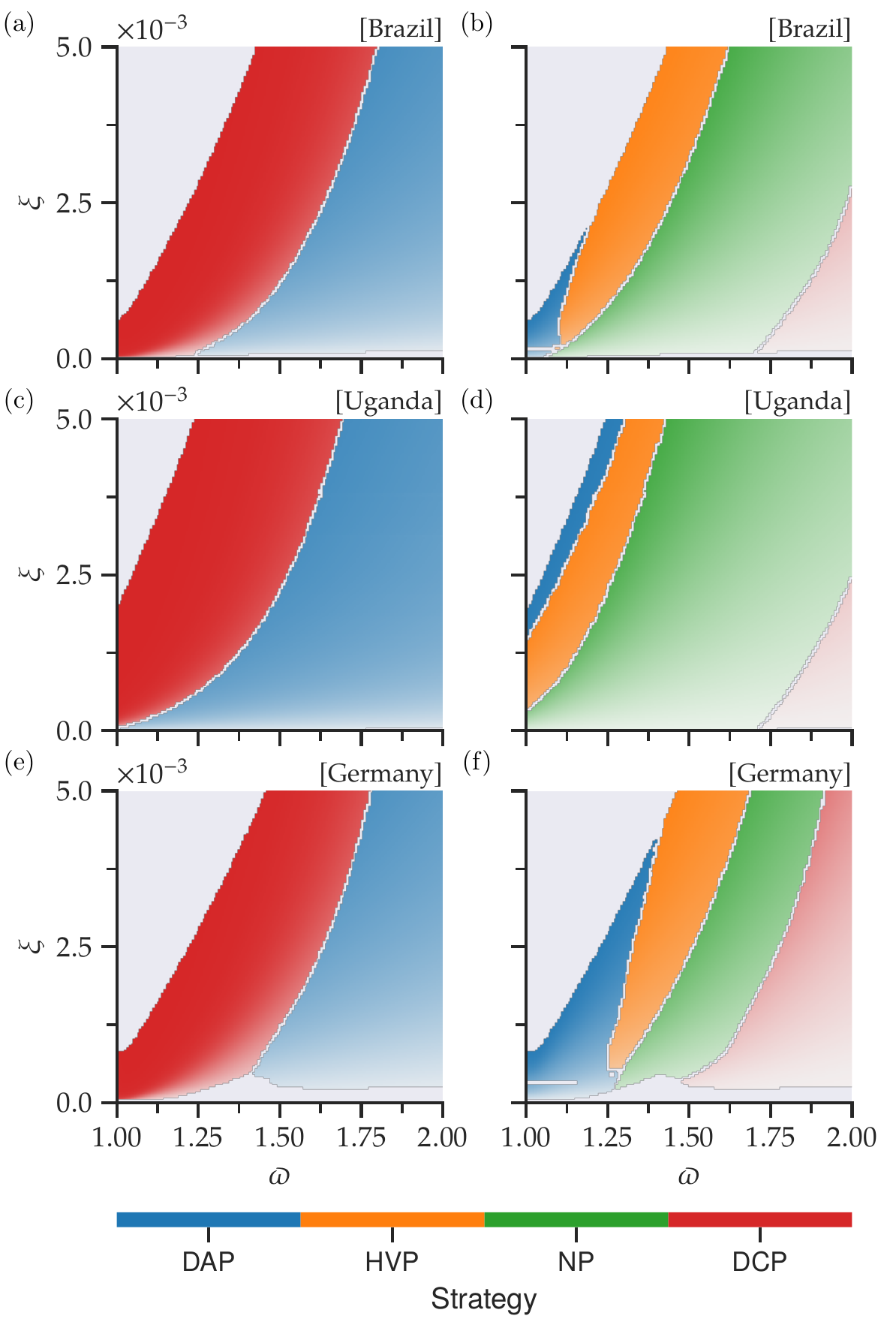}
\caption{{Comparison of the most and least effective strategies for different
demographic profiles.} Diagrams indicating the  optimal (left)   and  least effective (right) strategies in the parameter space   $\xi$ versus $\lbpar$  for (a,b) Brazil, (c,d) Uganda,  and (e,f) Germany. The COVID-19's IFR age profile, a time delay of $t_\mathrm{v}= 30$~days, vaccination with age-dependent values of efficacy against death and infection, and social distancing contact scenario were adopted. {Colors as in Fig.~\ref{fig:best_with_all}.}}
\label{fig:best_worst_countries}
\end{figure}

{Figure~\ref{fig:best_worst_countries} presents the results of most and least effective strategies to  reduce deaths considering age-dependent values of efficacy against death and infection and COVID-19's IFR age profile for Brazil, Uganda and Germany, while the results for uniform values of efficacy are shown in Figs.~\ref{fig:best_with_all}({e,f}) and Fig.~SI-5 of the \SM.}  The plot for optimal strategies has the same qualitative patterns for the three countries with age prioritization (DAP) being the most effective for higher infection regimes and contact prioritization (DCP) for lower infection and high vaccination rates. Quantitatively, Brazil and Germany present very similar diagrams  despite the substantially higher fraction of elderly  individuals in Germany. For  Uganda, with a mostly young population, the diagram region where DAP outperforms DCP is larger than in Brazil and Germany. The last result seems counter-intuitive at a first glance since prioritizing the most exposed is expected to be more effective in a population with very few elderly (3.3\%; See Table~\ref{tab:demography}).  However, the most exposed population in Uganda are the young while in Germany and Brazil, adults perform more contacts; See Figs.~\ref{fig:demographic}, SI-1, and SI-2 of the {\SM}. Since the COVID-19's IFR for the adult population is still much higher than for the young, DAP is also the best strategy for Uganda's contact pattern.
	
The least effective strategy provides more complex diagrams, depending strongly on the contact pattern and demography of the country. No priority strategy dominates the phase diagram for demographics of Brazil and Uganda. Prioritizing contacts is the least effective option for  uncontrolled dissemination, more evidently for Germany's demography. Prioritizing age is the least effective option in a small region of the diagrams consisting of concomitantly low transmission and vaccination rate in the case of Brazil, with a broader region for Germany; See Figs.~\ref{fig:best_worst_countries}(b,f), respectively. The diagrams for optimal and least effective strategies for  young, adults, and elderly individuals  also present nonlinear effects. Moreover, the diagrams for the whole population are ruled by deaths of elderly; See Figs.~SI-6 and SI-7 of the \SM.
	
\section{Conclusion}
\label{sec:conclusion}
	
The rise of a new, highly transmissible, and lethal infectious disease implies an enormous logistic challenge to minimize the damages and, especially in the case of viral pathogens, quick development and massive distribution of vaccines to the entire population is the most, maybe the only, viable option to mitigate the impacts of the disease. Moreover, the capability of the viruses in mutating and thus evading the protection conferred by either previous infections or vaccination imposes a constant concern about the optimal prioritization strategy  to be adopted in a realistic scenario of a limited supply of vaccine shots. We are nowadays witnessing a remarkable  success of massive vaccination to reduce the severe cases of COVID-19, wherever it has been adopted. 
	
The choice of the optimal prioritization strategy aiming at reducing the number of severe cases and consequently of deaths is far from being trivial due to the wide pool of relevant epidemiological parameters involved in the analysis.
{Healthcare authorities should be aware that there is not a unique optimal strategy that performs better in all situations due to the nonlinear complexity of the subject.} 	 To contribute to this problem, we investigated the role of social contact patterns (Fig.~\ref{fig:demographic}) and infection fatality ratio (Fig.~\ref{fig:fatality}) on vaccine prioritization strategies using an age-structured compartmental model and a data-driven approach, in which real epidemiological parameters are used as inputs to the numerical analysis. Prioritization of the most vulnerable population (with high risk of death)  and of the most exposed individuals (who perform more contacts) were compared with no prioritization. We report that vaccines, even with modest protection against infections, are very effective to reduce fatality irrespective of the strategy. For age prioritization, which corresponds to the most vulnerable population in several infectious diseases and particularly in COVID-19, the age profiles of deaths  are significantly altered while the infection profile changes comparatively little. Another important outcome of the simulations is that the effectiveness of the vaccination depends strongly on the contagion mitigation. Delays in starting vaccination imply the ineffectiveness of vaccines if the contagion is uncontrolled.
	
The optimal and least effective strategy to reduce deaths also depends on the epidemic scenario and IFR age profile. Vaccination of the most exposed population first is more effective than of the most vulnerable individuals when the epidemic is highly controlled with a low transmission rate. The prioritization of the most vulnerable population becomes the optimal approach for highly contagious scenarios. However, nonlinear dependence on the vaccination and contagion rates, depending on the IFR  profile, is observed. Comparing COVID-19 and seasonal influenza IFRs, we report that the region in the epidemiological parameter space, where  prioritizing vulnerable persons  is the most effective strategy, is broader for the former, despite the qualitative similarity between them. 
{This is in agreement with other results in the literature, such as the adoption of a DAP-like strategy for COVID-19~\cite{Goldstein2021,Fitzpatrick2021,Matrajt2020,Castro2021}, and a mix of DAP-like and DCP-like strategies for influenza~\cite{Fitzpatrick2021,Medlock2009}.}
{It is also important to notice the differences between Fig.~\ref{fig:best_with_all}(e), for a vaccine with uniform protection  across different age groups, and Fig.~\ref{fig:best_worst_countries}(a), considering age-dependent values of efficacy. The region in which the DCP  outperforms DAP strategy is larger in the latter, for which the effectiveness of the vaccine for elderly individuals is assumed to be lower and blocking the transmission is more efficient.}
Finally, the diagrams in the epidemiological space parameter reporting the optimal strategy depend little on demography and social contact profile when comparing data for Brazil, Germany, and Uganda, which present very distinct patterns. However, the least effective strategy is very sensitive to demography and contact matrices, revealing a complex dependence on epidemiological parameters. 
	
The data-driven analysis developed in this work raises important issues,  from the perspective of nonlinear dynamical systems, that may be underestimated in applied mathematical or statistical epidemiological modeling. This kind of modeling can help decision-makers to select the vaccination prioritization strategy according to the current scenario, but our central aim is to quantitatively address the importance of epidemiological parameters on the outcomes of the theoretical analysis using a mechanistic approach. Some simple, but still essential messages were presented. Beyond the obvious ones reporting that the faster and earlier the vaccination, the better its result is, we also show that the outcomes depend  nonlinearly on the epidemiological situation and particularities of the infectious disease. We expect that our more mechanistic approach can join statistical inference methods to provide more accurate responses to vaccination prioritization strategies.

\section*{Supplementary Material}

See \href{https://arxiv.org/src/2201.02869/anc/covid19-vac_supp.pdf}{supplementary material} for demography and contact structure of Uganda and Germany, impact of DAP vaccination for uniform values of efficacy, reduction of recovered individuals with age-dependent values of efficacy, optimal and least effective strategies for Uganda and Germany contact patterns, and optimal and least effective strategies for young, adult and elderly populations.
		
\begin{acknowledgments}
A.S., W.C., and S.C.F. acknowledge financial support from Coordenação de Aperfeiçoamento de Pessoal de Nível Superior -- CAPES https://www.gov.br/capes/ (grant No. 88887.507046/2020-00). S.C.F. acknowledge financial support from Conselho Nacional de Desenvolvimento Científico e Tecnológico -- CNPq https://www.gov.br/cnpq/ (grants No. 430768/2018-4 and 311183/2019-0) and Fundação de Amparo à Pesquisa do Estado de Minas Gerais -- FAPEMIG https://fapemig.br/ (grant No. APQ-02393-18). This study was financed in part by Coordenação de Aperfeiçoamento de Pessoal de Nível Superior -- CAPES - Finance Code 001. The funders had no role in study design, data collection and analysis, decision to publish, or preparation of the manuscript.
		\end{acknowledgments}
		
		\section*{Data Availability}
		
		The data that support the findings of this study are openly available in the Github repository at \url{https://github.com/wcota/covid19-vac-st/}.
		
		\section*{Conflict of Interest}
		
		The authors have no conflicts to disclose.
		
		\section*{Authors Contributions}
		
		\noindent{\bf Conceptualization, Supervision, Project Administration, and Original Draft Preparation:}  Wesley Cota, Silvio C. Ferreira; {\bf Data Curation and Software:} Arthur Schulenburg, Wesley Cota; {\bf Visualization:} Wesley Cota; {\bf Funding Acquisition:} Silvio C. Ferreira; {\bf Formal Analysis, Investigation, Methodology. Validation, and Review \& Editing:} Arthur Schulenburg, Wesley Cota, Guilherme S. Costa, Silvio C. Ferreira.

		\bibliography{refs.bib}

	\end{document}